\documentclass[12pt]{article}\usepackage{a4}
\newcommand{\be}{\begin{equation}}
\newcommand{\ee}{\end{equation}}
\newcommand{\bea}{\begin{eqnarray}}
\newcommand{\eea}{\end{eqnarray}}
\newcommand{\nn}{\nonumber}
\newcommand{\ep}{\epsilon}
\newcommand{\al}{\alpha}
\newcommand{\la}{\lambda}
\newcommand{\Ref}[1]{(\ref{#1})}
\newcommand{\D}{{\rm D}}
\newcommand{\m}{{\rm m}}
\usepackage{epsfig}
\begin{document}
\title{Groundstate projection of the charged SU(2) polarization tensor in a chromomagnetic background field}
\author{M.~Bordag\thanks{e-mail: Michael.Bordag@itp.uni-leipzig.de}\\
{\small University of Leipzig, Institute for Theoretical Physics,} \\
{\small  Postfach 100 920, 04009  Leipzig, Germany} \\ [12pt]
{\small and} \\ [12pt]
V. Skalozub\thanks{e-mail: Skalozubv@daad-alumni.de}\\
{\small Dnepropetrovsk National University, 49010 Dnepropetrovsk, Ukraine}}

\maketitle
\begin{abstract}
We consider the polarization tensor of a color charged gluon field in SU(2) in the background of a homogeneous color magnetic field and calculate its projection onto the lower (tachyonic) state. We obtain a quite simple representation of the polarization tensor in terms of double  parametric integral. We provide examples for the numerical evaluation of the polarization tensor in this case. Also, we consider the corresponding expressions at high temperature and calculate the charged gluon  magnetic mass in this limit. We conclude that radiation corrections  stabilize the ground state.  Applications of the results obtained are  discussed.
\end{abstract}
\section{Introduction}
Color magnetic fields are actively discussed in some areas of modern physics. So, the spontaneous generation of such fields prior to the electroweak phase transition is considered in \cite{Skalozub:1999bf,demc08-71-180,demc02-25-291,eliz11.1106.4962} as candidate source for the intergalactic magnetic fields observed today \cite{khar08-803-227,skok09-24-5925}. Strong  magnetic fields are considered to be created for a short time in heavy ion collisions at LHC \cite{deli10-82-051501,bali11-1111.4956}. For instance, their influence on the temperature of the de--confinement and chiral phase transitions is discussed.

Especially interesting quantities are the temperature induced masses which determine the characteristics of the quark-gluon plasma. Without magnetic field, the Debye mass $m_\D$ (defined as $m_\D^2=\Pi_{44}(T,k_4=0,\vec{k}\to 0)$ from the gluon polarization tensor) is of order $m_\D\sim gT$  (where $g$ is the gauge coupling constant and $T$ is the temperature) and the magnetic mass, defined in a similar way from the transversal components of the polarization tensor,  is of order $m_\m\sim g^2T$. In this way, both, chromo--electric and chromo--magnetic fields are screened in hot QCD. These masses were calculated by different methods, for instance analytically from the  1--loop polarization tensor  in \cite{kala84-32-525}, or by lattice calculations in \cite{boyd96-469-419}. We remind the corresponding results in QED \cite{land87-145-141}, resulting in $m_\D\sim eT$ with the electric charge $e$, and a zero magnetic mass, $m_\m=0$. As a consequence, the magnetic fields remains long range in hot QED.

In the presence of a color magnetic background field, these characteristics become modified. In this case it is reasonable to rewrite color gluon field, $V^a_\mu$  ($a = 1,2,3 $),  in terms of   charged, $ W_\mu^{\pm} = \frac{1}{\sqrt{2}}(V^1_\mu \pm i V^2_\mu),  $  and neutral, $V^3_\mu = A^3_\mu$, ones. The Debye mass becomes a function of the magnetic field, $m_\D=m_D(T=0)f(T/\sqrt{gB})$, where $f(T/\sqrt{gB})$ is some function turning into unity for $B\to0$, but reducing $m_\D$ for any finite $B$ (see \cite{bord07-75-125003} for the neutral  and \cite{bord08-77-105013} for the charged gluons).  For the magnetic mass a different picture is observed. For the color neutral gluon, in \cite{bord07-75-125003} the magnetic mass $m_\m$ was found to vanish, $m_\m=0$, and the account for polarization tensor results merely in a renormalization of the external field like in QED.
%% change Jan 4
%For the color charged gluon, the corresponding projection of the polarization tensor is still not calculated in a sufficiently explicit form. The present paper is devoted to closing this gap.
%Specifically, we start from the general representation derived in \cite{bord08-77-105013} and calculate explicitly its projection in the ground state. Thereby we pay special attention to the structure of the ultraviolet divergent pieces and confirm their independence on the background field.

It is interesting to note that the structure of the polarization tensor makes the characteristics of the gluon field at finite temperature in a magnetic background field quite different from those without the  field. If one considerers the polarization tensor  in one--loop approximation at finite temperature without background field, a fictitious pole appears \cite{kala84-32-525,klim81-33-934}. One needs to do re--summation (e.g., the hard thermal loop re--summation \cite{braa90-64-1338}) to get rid of it. The emerging masses are $m_\D\sim gT$ for the electric and $m_\m\sim g^2T$ for the magnetic components. In opposite, in a magnetic background field, there is no fictitious pole already at the one loop level at any temperature. As shown in \cite{bord07-75-125003}, the projection of the right hand side of the Schwinger-Dyson equation, $D^{-1}(k^2)=k^2-\Pi^2$, does not change sign. For example, for high $T$, the projection on the ground state takes the form $\langle D^{-1}(k^2)\rangle=k^2(1+5.7 \,T/\sqrt{gB})$. Hence, in the magnetic background no re--summation is necessary.

In the present paper we use the representation of the polarization tensor of the color charged gluons obtained in \cite{bord08-77-105013} and calculate the projection on the ground state. This is the tachyonic state on tree level. This state is given by
\be\label{ts1}\mid t\rangle_\mu  =\frac{1}{\sqrt{2}}
                                    \left(\begin{array}{c}1\\ i \\0\\0\end{array}\right)_\mu
                                    \mid 0 \rangle
\ee
(see eqs. (27) and (32) in \cite{bord06-45-159}).
An interesting question to answer here is whether the polarization tensor is able to remove the instability of this state.

%% changed Jan 4
For the color charged gluon, the corresponding projection of the polarization tensor is still not calculated in a sufficiently explicit form. The present paper is devoted to closing this gap.
 This case requires a special  consideration different from a general procedure  developed already in \cite{bord08-77-105013}. Thereby we pay special attention to the structure of the ultraviolet divergent pieces and confirm their independence of the background field. In addition,  the  high temperature case will be investigated.

%%%%%%%%%%%%%%%%%%%%%%%%%%%%%%%%%%%%%%%%%%%%
In the next section we adduce  the necessary formulas and notations for the polarization tensor. The actual calculation for the zero temperature  is carried out in the appendix. In sect.3 we consider the high temperature limit for the polarization tensor and calculate the magnetic mass. The conclusions and discussion are given in the last section.

%%%%%%%%%%%%%%%%%%%%%%%%%%%%%%%%%%%%%%%%%%%%%%%%%%%%%%%%%%%%%%%%%%%%%%%%%%%%%%%
\section{Calculation of the polarization tensor}
We use the representation of the polarization tensor given in \cite{bord08-77-105013}. In momentum representation, the initial expression reads
\bea\label{Pi}
\Pi_{\la\la'}(p)&=&\int\frac{dk}{(2\pi)^4} \
    \left\{\Gamma_{\la\nu\rho}G_{\nu\nu'}(p-k)\Gamma_{\la'\nu'\rho'}G_{\rho\rho'}(k)\right.
\nn \\ && \left.
    +(p-k)_{\la}G(p-k)k_{\la'}G(k)+k_{\la}G(p-k)(p-k)_{\la'}G(k)\right\}
\nn\\ [6pt]&&
    +\Pi^{\rm tadpol}_{\la\la'} \ ,
\eea
where the second line results from the ghost contribution
and the tadpole contribution is given by
\bea\label{Tp}
    \Pi^{\rm tadpol}_{\la\la'}&=&
    \int\frac{dk}{(2\pi)^4} \ \left\{\delta_{\la\la'}G_{\rho\rho}(k)-G(k)_{\la\la'}
\right\} \nn \\ &&
    +\int\frac{dp}{(2\pi)^4} \ \left\{
    \delta_{\la\la'}G_{\rho\rho}(p)+G_{\la'\la}(p)-2G_{\la\la'}(p)\right\}.
\eea
The vertex factor,
\be\label{Vf}
    \Gamma_{\la\nu\rho}=
    (k-2p)_\rho  \ \delta_{\la\nu}+\delta_{\rho\nu}(p-2k)_\la+\delta_{\rho\la}(p+k)_\nu \ ,
\ee
completes the description of the polarization tensor. These formulas hold also in a background field, provided the corresponding expressions for the propagators are used. We take a homogeneous magnetic background field in the representation where it is parallel to the third axes is both, configuration and color spaces,
\be\label{bf}   B_\mu^a=B \ \delta_{\mu 3}\delta^{a 3}.
\ee
After turning the momentum and the gluon field in color space into the charged basis,
\be\label{Bma}
p_\mu=B_{\mu\alpha}\ p_\alpha \quad \mbox{with} \quad B_{\mu\alpha}=\left(\begin{array}{cc}\frac{1}{\sqrt{2}}
\left(\begin{array}{cc}1&1\\i&-i\end{array}\right) & 0\\ 0 & \left(\begin{array}{cc}1&0\\0&1\end{array}\right) \end{array}\right)_{\mu\alpha} ,
\ee
we have a color neutral and a color charged field, both of spin 1. The charged field will occupy in the presence of the background field  Landau levels and we expand with respect to the corresponding eigenfunctions. In this space, the momentum $p_\mu$ of the charged gluon  is an operator, whose components fulfill the commutation relation
\be\label{com} [p_\al,p_\beta]=iF_{\al\beta}\equiv i\left(\begin{array}{cccc}
        -1&0&0&0\\0&1&0&0\\0&0&1&0\\0&0&0&1
        \end{array}\right)_{\al\beta}B.
\ee
In this expression we write $B$ instead $g B$. In what follows, where it will be not misleading we also  put $B = 1$,  for short.

A basis in this space is given by the vectors $\mid n,\sigma\rangle_\mu$, eq. (30) in \cite{bord06-45-159}. The tree level energies of these states are
\be\label{En}E_n=l_4^2+l_3^2+B(2n+1+2\sigma)\qquad(n=0,1,\dots, ~~\sigma=\pm 1),
\ee
where $l_3$ and $l_4$ are the momenta in parallel to the background field and imaginary time, respectively. The lowest state is $n=0$ with $\sigma=-1$ and it is called tachyonic for allowing for negative energy, $E_n<0$. In the charged basis, the lowest ({\it tachyonic}) state is
\be\label{t}
    \mid t\rangle_\al\equiv \mid 0,-1\rangle_\al
    =\left(\begin{array}{c}1\\0\\0\\0\end{array}\right)_\al\mid 0\rangle ,
\ee
where $\mid 0\rangle$ is the lowest Landau level which is annihilated by the operator $a$ in
\be\label{pal}p_\al=\left(\begin{array}{c}ia^\dagger\\-ia\\l_3\\l_4\end{array}\right)_\al
\ee
and we note
\be\label{pmual}p_\mu\mid t\rangle_\mu=p^\dagger_\al\mid t\rangle_\al=0.
\ee
Below, we will also use the notations $l^2=l_3^2+l_4^2$ and $h^2=p_1^2+p_2^2=aa^\dagger+a^\dagger a$.

Now, we continue with using the representation of the polarization tensor as given by eq. (51) in \cite{bord08-77-105013}. It results from the proper time representation of the propagators,
\be \label{st-repr} G(p - k) = \int\limits_0^{\infty} d s\ e^{- s (p - k)^2}, ~~~~~G( k) = \int\limits_0^{\infty} d t\ e^{- t  k^2},
\ee
and integration over $k$ in eq. \Ref{Pi}.
 First we consider the case $T=0$ and drop the sum over $N$ in the equation mentioned. The representation is in terms of a parametric  integral over the parameters $s$ and $t$,
\be\label{Pi1}
\Pi_{\la\la'}=
  \int_0^\infty ds\int_0^\infty dt \  {\Theta(s,t)}  \left( \sum_{i,j} {M}^{i,j}_{\la\la'}+
                                 {M}^{\rm gh}_{\la\la'}\right)
+ \ \Pi^{\rm tadpol}_{\la\la'}
\ee
with
\be\label{Pitp} \Theta(s,t)=\frac{\exp(-H)}{(4\pi)^2(s+t)\sqrt{\Delta}}.
\ee
Here the following notations are used:
\bea\label{nota}    H&=&\frac{st}{s+t}\, l^2+m(s,t)h^2\,,          \nn\\
                    m(s,t)&=&s+\frac12\ln\frac{\mu_-}{\mu_+}\,,   \nn\\
                    \Delta&=&\mu_- \, \mu_+ ,\nn\\
                    \mu_\pm&=&t+\sinh(s)e^{\pm s},
\eea
which are equivalent to eqs. (23-26) in \cite{bord08-77-105013}. The sum over $i,j$ in \Ref{Pi1} follows the subdivision introduced in \cite{bord08-77-105013} and the functions ${M}^{i,j}_{\la\la'}$ are given by eq. (53) in \cite{bord08-77-105013}.

Now we take the tachyonic projection of $\Pi_{\la\la'}$, eq. \Ref{Pi1}. In doing so we note especially $h^2=1 $ (for $B = 1)$ and the function $\Theta$ simplifies,
\be\label{Th} \Theta(s,t)_{|h^2=1}=\frac{\exp(-\frac{st}{s+t}l^2-s)}{(4\pi)^2(s+t)\mu_-}.
\ee
For the projection of the functions ${M}^{i,j}_{\la\la'}$ we use representation (55) in \cite{bord08-77-105013}. We collect them in the appendix, together with the contributions from the tadpole graphs. At this place we mention that under the tachyonic projection we get directly a representation suitable for further calculations. This is in opposite to the general case in \cite{bord08-77-105013}, where we were forced to integrate by parts in a special way in order to account for the transversal structure of the polarization tensor already on the level of the parametric integrals.

From eq. \Ref{M7} in the appendix we get the non-regularized expression. Adding a dimensional  regularization ($\ep>0$ with the renormalization mass scale $\mu$) and restoring the dependence on the magnetic field $B$ we get
\bea\label{Pireg}    \langle t\mid \Pi\mid t\rangle &=&
        \frac{B}{(4\pi)^2} \Bigg\{
            \int\frac{ds\, dt}{s+t} \left(\frac{\mu^2 q}{B}\right)^\ep
            \left(\frac{4}{\mu_-}+4\frac{s+te^{2s}}{s+t}\frac{l^2}{B}\right)\ \frac{e^{-\frac{st}{s+t}\frac{l^2}{B}-s}}{\mu_-}
\nn\\[5pt]&&
        +\int_0^\infty\frac{dq}{q}    \left(\frac{\mu^2 q}{B}\right)^\ep
                \left(\frac{-2}{q}-\frac{2}{\sinh(q)}-4\cosh(q)\right)    \Bigg\}.
\eea
Using variables 
\be\label{st}   s=qu,~~~~t=q(1-u),
\ee
it is easily seen that the linear ultraviolet divergencies, which manifests themselves in the behavior for $q\to0$,  cancel (this is a consequence of the gauge transformation properties of the polarization tensor \cite{Bordag:2005zz}).  The logarithmic ones remain and these take the form
\be\label{Pidiv1}   \langle t\mid \Pi\mid t\rangle =
                 \frac{1}{(4\pi)^2}
                 \int_0 \frac{dq}{q}
                 \left(\frac{\mu^2 q}{B}\right)^\ep
                 \left(\frac{10}{3}(l^2-B)+O(1)\right).
\ee
This is just the same as in case with zero magnetic background field (keeping, however, the basis of Landau level states). Hence we can use the standard renormalization prescription as to subtract the polarization tensor with zero field. Specifically we define
\bea\label{Pidiv2} \langle t\mid \Pi\mid t\rangle^{\rm div} &\equiv &
 \frac{1}{(4\pi)^2}
                 \int_0^\infty\frac{dq}{q}
                 \left(\frac{\mu^2 q}{B}\right)^\ep
                  \frac{10}{3}\ \frac{l^2-B}{1+q}.
\nn\\   &=&  \frac{1}{(4\pi)^2} \left(\frac{1}{\ep}+\ln\frac{\mu^2}{B} \right)
                 \frac{10}{3}\ (l^2-B)+O(1).
\eea
The factor $1/(1+q)$ was introduced for convenience. This is an alternative to an auxiliary gluon mass. The arbitrariness brought in by this factor corresponds to a finite renormalization or a change in the parameter $\mu$.

Now we define the renormalized polarization tensor by
\be\label{Piren0}   \langle t\mid \Pi\mid t\rangle^{\rm ren}  =
        \langle t\mid \Pi\mid t\rangle  -\langle t\mid \Pi\mid t\rangle^{\rm div} ~_{|\ep=0}.
\ee
Using \Ref{Pireg}, \Ref{Pidiv2} and the variables \Ref{st}, it takes the form
\bea\label{Piren1} \langle t\mid \Pi\mid t\rangle^{\rm ren}  &=&
    \frac{B}{(4\pi)^2} \int_0^\infty  \frac{dq}{q}
        \Bigg\{   
 \nn\\&&      \int_0^1du\,\frac{q}{\mu_-}\left(\frac{4}{\mu_-}+4 (u +(1 - u) e^{2 q u }) \frac{l^2}{B}\right)
                          \frac{  e^{-q u (1 - u)\frac{l^2}{B}- q u}}{\mu_-}
\nn\\&& -\frac{2}{q}-\frac{2}{\sinh(q)}-4\cosh(q)-\frac{10}{3}\frac{\frac{l^2}{B}-1}{1+q} \Bigg\}.
\eea
This expression is ultraviolet finite. However, it has still a divergence resulting from the tachyonic mode in the loop integration. This divergence manifests itself in the $s$-integration for $s\to\infty$. To understand this divergence one needs to remember that we already performed the Wick rotation. The initial expression needs to be written in Minkowski space with the corresponding $'\!+i\ep\,'$-prescriptions in the propagators. Starting from there, one needs to identify a contribution involving the poles (in momentum plane) from the tachyonic mode in the loop integration ({\it tachyonic part}). That contribution may be anti-Wick rotated (in order to keep convergence) which is $q\to -i q$ in the parametric representation. The remaining part may be Wick rotated as usual ($q\to iq$). We mention that the identification of the tachyonic part is not unique which, however,  does not influence the final result. In representation \Ref{Piren1}, once we already Wick rotated the whole expression, we can identify the tachyonic part and anti-Wick rotate it twice, substituting $q\to qe^{i\pi}$.

The tachyonic part is just  that which, after Wick rotation is exponentially divergent for $s\to\infty$. In order to identify this part we represent \Ref{Piren1} in the form
\be\label{Pit1} \langle t\mid \Pi\mid t\rangle^{\rm ren} =A_1+A_2+B_1+B_2,
\ee
where we defined from the double integral
\be\label{A1}A_1=\frac{1}{(4\pi)^2}\int\frac{ds\, dt}{s+t}\
        \frac{te^{s}}{s+t}\ l^2 \ \frac{e^{-\frac{st}{s+t}\frac{l^2}{B}}}{t+\frac12}
\ee
and, from the single integral,
\be\label{A2} A_2=\frac{B}{(4\pi)^2}\int_0^\infty\frac{dq}{q}
        \left(\frac{\mu^2q}{B}\right)^\ep   (-2) \ e^q \, .
\ee
These are the tachyonic parts, and $B_1$ and $B_2$ collect the remaining contributions, see below.

We mention that $A_1$ has no ultraviolet divergence and we could put $\ep=0$ in \Ref{A1}. Now we perform in these two expressions the anti-Wick rotation. We get
\be\label{A1r} A_1=\frac{1}{(4\pi)^2}\int\frac{ds\, dt}{s+t}\
        \frac{te^{-s}}{s+t}\ l^2 \ \frac{e^{ \frac{st}{s+t}\frac{l^2}{B}}}{t-\frac12-i0},
\ee
where the $'\!-i0\,'$-prescription results from turning the integration path. It makes $A_1$ having an imaginary part. Now $A_1$, eq. \Ref{A1r}, is a finite expression for $l^2<B$ and can be calculated directly, numerically for instance.

The part $A_2$, after anti-Wick rotation, becomes
\be\label{A2r} A_2=\frac{B}{(4\pi)^2}\int_0^\infty\frac{dq}{q}
        \left(\frac{\mu^2q \,e^{i\pi}}{B}\right)^\ep   (-2) \ e^{-q}
\ee
and can be calculated explicitly,
\be\label{A2r1}A_2=A_2^{\rm ren}+A_2^{\rm div},
\ee
with
\be\label{A2r2}A_2^{\rm div}=-2 \frac{B}{(4\pi)^2} \left(\frac{1}{\ep}+\ln\frac{\mu^2}{B}\right)
\ee
and
\be\label{A2r3}A_2^{\rm ren}=-2\pi i,
\ee
where we made the subdivision according to the ultraviolet divergence in $A_2$.

\begin{figure}[h]
\includegraphics[width=11cm]{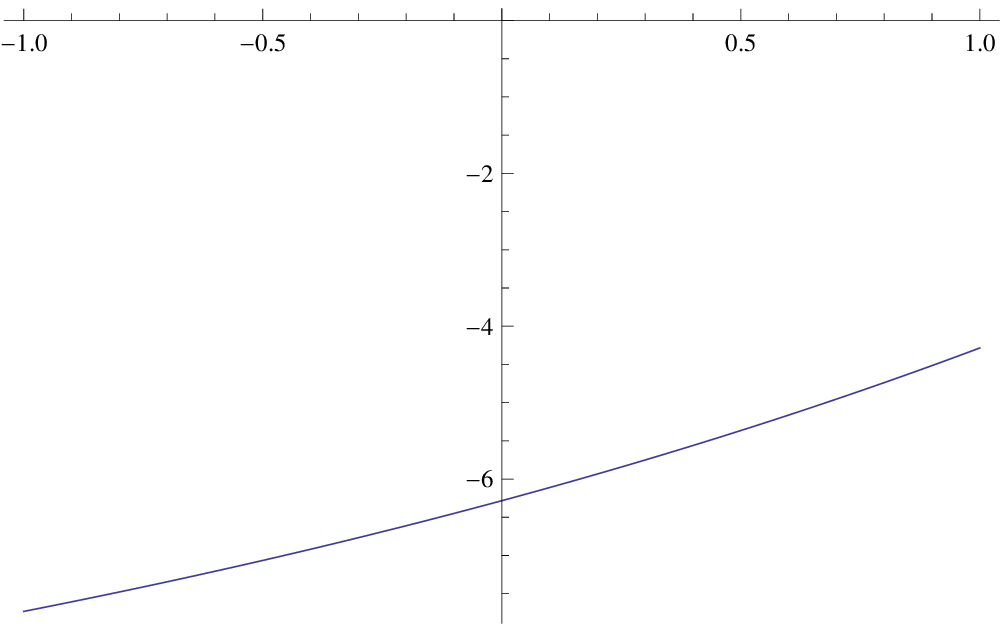}\\[10pt]
\caption{The function $f_{\rm im}$   in eq. \Ref{Pire} as function of its argument.}\label{fig1b}
\end{figure}
\begin{figure}[h]
\includegraphics[width=11cm]{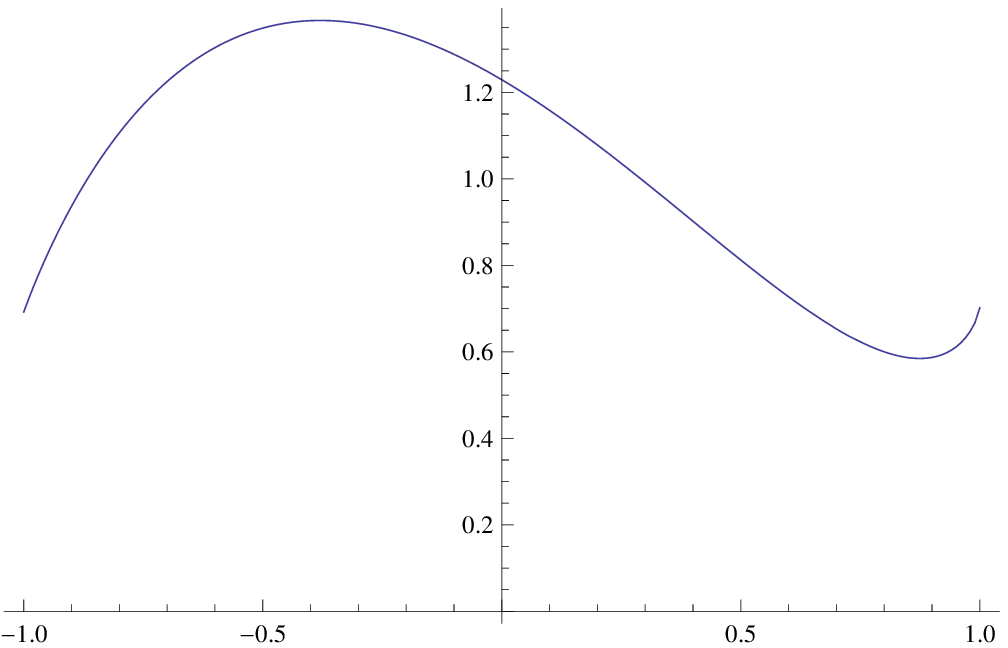}
\caption{The function   $f_{\rm re}$  in eq. \Ref{Pire} as function of its argument.}\label{fig1a}
\end{figure}

Finally we have to calculate $B_1$ and $B_2$. We represent them in the form
\be\label{B1} B_1=B_1^{\rm ren}+B_1^{\rm div},
\ee
with
\bea\label{B1a}  B_1^{\rm ren}  &=&
        \frac{B}{(4\pi)^2}
            \int\frac{ds\, dt}{s+t} \Bigg\{%\left(\frac{\mu^2 q}{B}\right)^\ep
            \left(\frac{4}{\mu_-}+4\frac{s+te^{2s}}{s+t}
                -4\frac{te^{-s}}{s+t}\frac{l^2}{B}
            \right)\ \frac{e^{-\frac{st}{s+t}\frac{l^2}{B}-s}}{\mu_-}
\nn\\[5pt]&&~~~~~~~~~~~~~~~~~~~~
         -\frac{4}{s + t}-4\frac{s(s - t)+(s^2 + t^2 + s t)\frac{l^2}{B}}{(s+t)^3(1+ s + t)}
         \Bigg\}
\eea
and
\be\label{B1b}B_1^{\rm div}=\frac{B}{(4\pi)^2}
            \int\frac{ds\, dt}{s+t}  \left(\frac{\mu^2q  }{B}\right)^\ep
            \left[  \frac{4}{s + t}+4\frac{s(s - t)+(s^2 + t^2 + s t)\frac{l^2}{B}}{(s+t)^3(1+ s + t)}\right]
\ee
for the part\ resulting from the double integral and
\be\label{B2}B_2=B_2^{\rm ren}+B_2^{\rm div},
\ee
with
\bea\label{B2a}B_2^{\rm ren}&=&\frac{B}{(4\pi)^2}
        \int_0^\infty\frac{dq}{q}\left(\frac{-2}{q}-2e^{-q}-
            \frac{2}{\sinh(q)}+\frac{4}{q}+\frac{2}{1+q}\right)
\nn\\&\simeq&    2.54 \frac{B}{(4\pi)^2}
\eea
and
\be\label{B2b}B_2^{\rm div}=\frac{B}{(4\pi)^2}
         \int_0^\infty\frac{dq}{q}  \left(\frac{\mu^2q  }{B}\right)^\ep
         \left(\frac{-4}{q}-\frac{2}{1+q}\right)
\ee
for the contribution from the integral over $q$. The divergent contributions collect into \Ref{Pidiv2},
\be\label{Pidiv3}\langle t\mid \Pi\mid t\rangle^{\rm div}=A_2^{\rm div}+B_1^{\rm div}+B_2^{\rm div},
\ee
and undergo the renormalization.

In this way, we get for the renormalized polarization tensor
\bea\label{Pire}
    Re(\langle t\mid \Pi\mid t\rangle^{\rm ren})&\equiv&
        \frac{B}{(4\pi)^2}\ f_{\rm re}\left(\frac{l^2}{B}\right)
        = Re(A_1)+B_1^{\rm ren}+B_2^{\rm ren},
\nn\\
   Im(\langle t\mid \Pi\mid t\rangle^{\rm ren})&\equiv&
        \frac{B}{(4\pi)^2}\ f_{\rm im}\left(\frac{l^2}{B}\right)
        = Im(A_1)+Im(A_2^{\rm ren}).
\eea
The functions $f_{\rm re}$ and $f_{\rm im}$ are shown in Fig.1.

\section{Magnetic mass at high temperature}
The above formulas were obtained for zero temperature, $T=0$. These allow for immediately writing down the corresponding results for high temperature, $T\to\infty$. To do that, first one has to return   to eq. \Ref{Pi}. At finite temperature $T$, in that formula, the integration over the momentum $k_4$ must be substituted by the Matsubara sum,
\be\label{sum} \int\limits_{-\infty}^{\infty} \frac{d k_4}{2 \pi} \to
    T \sum\limits_{l = -\infty}^{\infty}
\ee
in the otherwise unchanged expression. From here, the limit of high temperature, $T\to\infty$, is obtained by taking the contribution of the zeroth Matsubara frequency, $l=0$. This is, of course, the known dimensional reduction to a theory without temperature in three dimensions. In the representation in terms of parametric integrals, eq. \Ref{Pireg}, the integrand is simply 'missing' a factor $\sqrt{4\pi q}$. Hence, the tachyonic projection of the polarization tensor \Ref{Pireg} takes the form
\bea\label{PiT}    \langle t\mid \Pi\mid t\rangle &\raisebox{-5pt}{${\sim\atop T\to\infty}$}&
        \frac{T\sqrt{B}}{(4\pi)^{3/2}} \Bigg\{
            \int\frac{ds\, dt}{\sqrt{s+t}}
            \left(\frac{4}{\mu_-}+4\frac{s+te^{2s}}{s+t}\frac{l^2}{B}\right)\ \frac{e^{-\frac{st}{s+t}\frac{l^2}{B}-s}}{\mu_-}
\nn\\[5pt]&&
        +\int_0^\infty\frac{dq}{\sqrt{q}}
                \left(\frac{-2}{q}-\frac{2}{\sinh(q)}-4\cosh(q)\right)    \Bigg\},
\eea
where now $l^2=l_3^2$. In this expression, we removed the regularization. This is possible since the \Ref{PiT} does not contain ultraviolet divergencies. The former linear ones cancel as before  (in four dimensions) and the former logarithmic ones do not appear due to the dimensionality.

In the following we focus on the magnetic mass of the charged gluon,
\be\label{mch}
                    m_{\rm ch}^2 = < t| \Pi(B, T, p_4 = 0, p_3 \to 0)| t>.
\ee
It is given by eq. \Ref{PiT} with $l^2=0$ in the tachyonic projection. So it remains to treat the tachyonic mode in \Ref{PiT}. Because of $l^2=0$  this is here simpler than in the preceding section.The tachyonic contribution results from $A_2$  \Ref{A2} only.  Accounting for the changes dimensionality it can be calculated easily and it delivers the imaginary part,
\be
    Im(\langle t\mid \Pi\mid t\rangle_{T\to\infty})=-2i\sqrt{\pi}\frac{T\sqrt{B}}{(4\pi)^{3/2}},
\ee
The real part comes completely from the $B_2$-part. It is also much simpler than in the preceding section and it reads
\bea\label{mmass}
   Re( \langle t\mid \Pi\mid t\rangle_{T\to\infty})&=&\frac{T\sqrt{B}}{(4\pi)^{3/2}}
    \int_0^\infty\frac{dq}{\sqrt{q}}
    \left[  \int_0^1 du\ \frac{4qe^{- q u}}{\mu_-^2}-\frac{2}{q}-\frac{2}{\sinh(q)}-2e^{-q}\right],
\nn\\&\simeq&-3.26 \frac{T\sqrt{B}}{(4\pi)^{3/2}}.
\eea
The last line is a result of numerical integration.

The main result of this calculation is that the projection of the polarization tensor is proportional to $T \sqrt{B}$ allowing for the conclusion that there is a field depentent magnetic mass. Further, it is important that the real part has negative sign. Hence it follows that the radiation corrections act to stabilize the spectrum of charged gluons at high temperature.
\section{Conclusions}
In the above sections we calculated the tachyonic projection of the polarization tensor of color charged gluon in a chromomagnetic background field. We used the general formulas derived earlier in \cite{bord08-77-105013}. The expression for the polarization tensor is given on the level of the Feynman rules by eqs. \Ref{Pi} and \Ref{Pitp}. We account for the magnetic background field by the known representation in terms of parametric integrals after  Schwinger. From simply inserting these formulas one comes to an expression where, already in QED, the transversal structure cannot be seen directly. Instead, one needs to integrate by parts in the parametric integrals to make this structure manifest in the integrands. A check for this procedure is the cancelation of the linear ultraviolet divergences. In the non-Abelian case the picture is more involved since the polarization tensor is not transversal as shown in \cite{Bordag:2005zz}. Nevertheless, in \cite{bord08-77-105013} a way was found how to perform the necessary partial integrations in order to make the corresponding structures, formulated in terms of form  factors, manifest.

In this way, in \cite{bord08-77-105013}, the general structure of the polarization tensor in terms of the parametric integrals was found. However, as it turned out, this structure does not hold in general. For the projection on the tachyonic state, which we consider in the present paper, the integration by parts does not work. Doing it formally, artificial divergences are introduced which manifest themselves as factors $\mu_+$ in the denominators which in fact should never appear.

For this reason we made in the present paper  a direct projection of the polarization tensor instead integration by parts. As a result we obtained the representation \Ref{Pireg}. Here the linear ultraviolet divergence cancels between the different contributing graphs (the basis graph, the ghosts and the tadpole graphs). We consider this a check for having the correct    properties with respect to the gauge symmetry.

The remaining calculation involves the anti-Wick rotation and the  renormalization according to the known rules. The result is quite simple and numerical examples are shown in Figs. 1 and 2.

On the base of these  calculations, we  found the charged gluon magnetic mass in the high temperature limit.   The derived result in eq. \Ref{mmass} is signalling that  radiation corrections act to stabilize the tree-level spectrum \Ref{En}. Really, if one considers  the pole of the Schwinger-Dyson operator equation, taken in the ground state, $< t|D |t>^{- 1} = < t |l^2 - gB  - \Pi( T, B, p_4 = 0, p_3 \to 0)| t > $, then the positivity of the  "effective mass squared" $m_{eff.}^2 = - g B + 3.26 \frac{T\sqrt{g B}}{(4\pi)^{3/2}}$  at high temperature follows. This result fills in the gap existed in calculations of the screening parameters for charged gluons. Due to spectrum stability, the stabilization of the Abelian chromomagnetic fields of the type in eq.\Ref{bf} at high temperature is guaranteed.

These results can find a number of applications in different fields. At finite temperature, these are the early Universe and heavy ion collisions where different types of  magnetic fields are expected to be generated. At zero temperature, the obtained results may serve  as starting point for the calculation of various characteristics of charged vector particles with gyromagnetic ratio $\gamma = 2$. An example is  the $\rho$-meson electrodynamics in strong magnetic fields, where  interesting properties of the vacuum were discussed in the literature (see, for example, \cite{djuk05-95-012001}, \cite{cher11-106-142003}, \cite{cher10-82-085011} and references therein). Another obvious candidate for application is a $W$-boson in magnetic fields. In the two latter cases, one needs to account for   the mass term of the particle, which can be incorporated  in the developed formalism easily.

\section*{Acknowledgement}
The authors benefited from exchange of ideas by the European Science
Foundation activity {\it New Trends and Applications of
Casimir Effect}.\\
One of us (VS) was supported by the mentioned network.

%%%%%%%%%%%%%%%%%%%%%%%%%%%%%%%%%%%%%%%%%%%%%%%%%%%%%%%%%%%%%%%%%%%%%%%%%%%%%%%
\section*{Appendix}
In this appendix we perform the projection of the functions $M^{ij}_{\la\la'}$, eq. \Ref{Pi1}, using the expressions given in  eq. (55) in \cite{bord08-77-105013}. We repeat that formulas here putting $\tilde{u}=0$ and stripping off a factor $Z$ since we do not need it in the current calculation. In \cite{bord08-77-105013} it was introduced for use in integrating by parts which we do not need to do here. Furthermore, we drop all contributions containing $p_\la$ in view of eq. \Ref{pmual} and obtain
\bea\label{M1}    M^{11}_{\la\la'}    &=& 2\left(\frac{2iF}{D^\top}\right)_{\la\la'}\ {\rm tr}E \, ,
\nn\\           M^{12}_{\la\la'}    &=&M^{21}_{\la\la'}    =0\, ,
\nn\\           M^{13_1}_{\la\la'}    &=&M^{3_11}_{\la\la'}    = - \left(E^\top\frac{2iF}{D^\top}\right)_{\la\la'},
\nn\\           M^{13_2}_{\la\la'}    &=&M^{3_21}_{\la\la'}    = - \left(E \frac{2iF}{D^\top}\right)_{\la\la'}\, ,
\nn\\           M^{22}_{\la\la'}    &=& 4S_{\la\la'}\left(pT^\top p\right)
                                        +4T^\top_{\la\la'}\left(pS p\right)
                                        +4ST^\top 2iF\, ,
\nn\\           M^{23_1}_{\la\la'}    &=&M^{3_12}_{\la\la'}    =
            - 2t\delta_{\la\la'}\left( p\frac{2iF}{D^\top}p\right)
            - \left( 2itF \frac{2iF}{D^\top}\right)_{\la\la'}\, ,
\nn\\           M^{23_2}_{\la\la'}    &=&M^{3_22}_{\la\la'}    =
            - 2E_{\la\la'}\left( p\left(\frac{A}{D}\right)^\top p\right)
             +\left( A \frac{2iF}{D^\top}\right)_{\la\la'}\, .
\eea
Here   notations from  \cite{bord08-77-105013} are used, which we rewrite somehow,
\bea\label{repr2}   \frac{2iF}{D^\top}  &=&
            \frac{\delta_{||}}{s+t}+\left(\frac{-1}{\mu_-}+\frac{1}{\mu_+}\right)iF
                                    +\left(\frac{1}{\mu_-}+\frac{1}{\mu_+}\right)\delta_\perp \,,
\nn \\
        \frac{A}{D}  &=&
            \frac{s}{s+t}\, \delta_{||}+\left(\frac{1-\al}{4\mu_-} -\frac{1-\al}{4\al\mu_+}\right)iF
                                    +\left(\frac{1-\al}{4\mu_-} +\frac{1-a\l}{4\al\mu_+}\right)\delta_\perp\,,
\nn \\
        S  &=&
              \delta_{||}+\left(-\frac{\al \mu_+}{\mu_-} -\frac{\mu_-}{\al\mu_+}\right)iF
              +\left(\frac{\al \mu_+}{\mu_-} -\frac{\mu_-}{ \al\mu_+}\right)\delta_\perp\,,
\nn \\
        T  &=&
              \delta_{||}+\left(-\frac{  \mu_+}{\mu_-} -\frac{\mu_-}{ \mu_+}\right)iF
              +\left(\frac{ \mu_+}{\mu_-} -\frac{\mu_-}{ \mu_+}\right)\delta_\perp\,,
\eea
and we introduced the new notations,
\be\label{almu} \al=e^{-2s},\qquad \mu_\pm=t+\sinh(s)\, e^{\pm s},
\ee
and we remind eq. (20) in  \cite{bord08-77-105013},
\be\label{40} E= \delta_{||}-\sinh(2s)iF+\cosh(2s) \delta_{\perp}.
\ee
In order to calculate the contributions quadratic in the momenta we note the formula
\be\label{41}\left( p(\mu \delta_{||}+\sigma iF+\nu \delta_{\perp})p\right)
            =\mu l^2-\sigma+\nu,
\ee
which holds for any $\mu$, $\sigma$, $\nu$ in the tachyonic state ($h^2=1$). From here we calculate
\bea\label{pp}\left( p\frac{2iF}{D^\top}p\right)  &=& \frac{l^2}{s+t}+\frac{1}{\mu_-}\,,
\nn \\  \left( p\left(\frac{A}{D}\right)^\top p\right)&=&
                \frac{s}{s+t}\,l^2+\frac{1-\al}{2\mu_-}\,,
\nn\\       \left(pT^\top p\right)  &=& l^2+\frac{\mu_+}{\mu_-}\,,
\nn\\       \left(pS p\right)     &=& l^2+\frac{\al\mu_+}{\mu_-}\,.
\eea
Further we need the tachyonic projection,
\be\label{tuv}\langle t\mid \mu \delta_{||}+\sigma iF+\nu \delta_{\perp}  \mid t\rangle
                =-\sigma+\nu\,,
\ee
which follows directly from eqs. (29) and (32) in  \cite{bord08-77-105013}. Now, using \Ref{pp} and \Ref{tuv}, we can calculate the averages in \Ref{M1},
\bea\label{M2}  \langle t\mid M^{11}  \mid t\rangle &=&4\frac{1+\cosh(2s)}{\mu_-},
\nn\\    \langle t\mid M^{13_1}  \mid t\rangle + \langle t\mid M^{3_11}  \mid t\rangle &=&
                            -\frac{2\al}{\mu_-},
\nn\\    \langle t\mid M^{13_2}  \mid t\rangle + \langle t\mid M^{3_21}  \mid t\rangle &=&
                            -\frac{2}{\al\mu_-},
\nn\\    \langle t\mid M^{22}  \mid t\rangle   &=&
                            4\left(\frac{1}{\al }+1\right)l^2,
\nn\\    \langle t\mid M^{23_1}  \mid t\rangle + \langle t\mid M^{3_12}  \mid t\rangle &=&
                            -\frac{4t}{s+t}\ l^2,
\nn\\    \langle t\mid M^{23_2}  \mid t\rangle + \langle t\mid M^{3_22}  \mid t\rangle &=&
                            -\frac{4s}{s+t}\ \frac{l^2}{\al}.
\eea
In these formulas some simplifications occurred. Now we add these contributions,
\be\label{M3} \langle t\mid \sum_{ij}M^{ij}  \mid t\rangle
        =\frac{4}{\mu_-}-4\frac{t+\frac{s}{\al}}{s+t}\ l^2,
\ee
where, again, a number of contributions canceled. It remains to consider the contribution from $M^{33}$ and from the ghosts (last line in eq. (55) in \cite{bord08-77-105013}). These combine into derivatives, see eq. (90) in \cite{bord08-77-105013}, which allows for carrying out one of the parameter integrations. Using
\be\label{46}\Theta(s=0,t)=\frac{1}{t^2},\qquad\Theta(s,t=0)=\frac{1}{s\sinh(s)},
\ee
we get in the projection
\be\label{M4}\int ds\, dt\ \langle t\mid \left(M^{33} +M^{\rm gh}\right)\Theta(s,t) \mid t\rangle =
\int\frac{dq}{q}\left(\frac{1}{q}+\frac{1}{\al\sinh(q)}\right).
\ee
Finally, we need the contributions from the tadpoles, \Ref{Tp}, which take the form
\be\label{M5} \langle t\mid\Pi^{\rm tp} \mid t\rangle =
\int\frac{dq}{q}\left(\frac{-3}{q}-\frac{2+\cosh(2qs)-3\sinh(2q)}{ \sinh(q)}\right).
\ee
Together,  \Ref{M4} and \Ref{M5}  collect  into
\be\label{M6}\langle t\mid \Pi^{33} +\Pi^{\rm gh}+\Pi^{\rm tp} \mid t\rangle
        =\frac{1}{(4\pi)^2}\int\frac{dq}{q}\left(\frac{-2}{q}-2\frac{1+\sinh(2s)}{\sinh(s)}\right).
\ee
Collecting from \Ref{M3} and \Ref{M6} we come to
\bea\label{M7}\langle t\mid \Pi  \mid t\rangle    &=&
    \frac{1}{(4\pi)^2}  \Bigg\{   \int\frac{ds\,dt}{s+t}\left[\frac{4}{\mu_-}+4\frac{s+\frac{t}{\al}}{\sinh(s)}\right]
    \ \frac{e^{-\frac{st}{s+t}l^2-s}}{\mu_-}
    \nn \\       &&  +\int\frac{dq}{q}\left(\frac{-2}{q}-\frac{2}{\sinh(s)}-4\cosh(q)\right)
    \Bigg\},
\eea
which represents the non-renormalized polarization tensor of the charged gluon in the projection to the lowest, the tachyonic state.
%%
%\be\label{}
%\ee
%%
%%
%\be\label{}
%\ee
%%

\bibliography{C:/Users/bordag/WORK/Literatur/bib/papers,C:/Users/bordag/WORK/Literatur/Bordag,C:/Users/bordag/WORK/Literatur/libri,lit}
\bibliographystyle{unsrt}

\end{document}